# Econometric Approach to Analyzing Determinants of Sustained Prosperity*

### Anika Dixit

## ABSTRACT

Every year, substantial resources are allocated to foreign aid with the aim of catalyzing prosperity and development in recipient countries. The diverse body of research on the relationship between aid and gross domestic product (GDP) has yielded varying results, finding evidence of both positive, negative, and negligible associations between the two.

This study employs econometric techniques, namely Fully Modified Ordinary Least Squares Regression (FMOLS) and the Generalized Method of Moments (GMM), to explore the intricate links between innovation and different types of official development assistance (ODA) with the overarching construct of prosperity. The paper also reviews the linkages between foundational metrics, such as the rule of law, education, and economic infrastructure and services, in enabling self-sustaining prosperity.

Drawing upon panel data of relevant determinants for 74 countries across the years 2013 to 2021, the study found that there was a negligible relationship between both ODA and innovation indices with prosperity. Notably, foreign aid targeted specifically toward education was observed to have a positive impact on prosperity, as was the presence of rule of law in a state.

The results of the study are then examined through the lens of a case-study on Reliance Jio, exemplifying how the company engineered an ecosystem that harnessed resources and facilitated infrastructure development, thereby contributing to self-sustaining economic growth and prosperity in India.

## 1. INTRODUCTION

Foreign aid constitutes a substantial allocation of resources annually, intended to stimulate prosperity and development in recipient countries. However, the precise impact of this economic assistance on the overall prosperity of nations—to be distinguished from their GDP growth— remains a complex and multifaceted issue.

In this context, the measurement and definition of prosperity encompass a broad spectrum of economic, social, and developmental factors and is assessed through the Legatum Prosperity Index (**1**).

Over the past six decades, the annual disbursement of aid has increased more than fourfold, surging from US$38 billion in 1960 to US$204.0 billion in 2022 (**2**). While global aid has supported diverse projects across sectors—from fulfilling basic needs for individuals living in poverty to providing humanitarian aid, health services, and emergency food assistance—it has not been effective in facilitating sustainable, long-term growth and prosperity.

Despite these efforts, widespread poverty continues to persist worldwide, with limited prospects for significant improvement in the near future. Nearly half of the global population subsists on less than US$6.85 per person per day, with projections suggesting that nearly 600 million individuals will live on less than US$2.15 per day by 2030 (**3**).

Nevertheless, there is optimism that nations can uplift themselves from poverty through the strategic deployment of innovation and entrepreneurship as catalysts for sustainable economic growth. This study seeks to substantiate this notion by examining the hypothesis that Official Development Assistance (ODA) or aid alone may not be sufficient to significantly advance lasting prosperity. Instead, it postulates that a dynamic framework driven by innovation and entrepreneurship, facilitated by education and the rule of law, is essential for fostering sustained economic progress and prosperity.

## 2. REVIEW OF LITERATURE

Over the past decade, there has been considerable review of the relationship between foreign aid and the gross domestic product (GDP) of recipient nations. This paper, however, instead assesses development under the lens of the Legatum Prosperity Index; unlike GDP, which fails to account for non-monetary items that add value to an economy, the Legatum Index offers a more holistic review of development within a nation, encompassing factors such as education and rule of law. There is limited pre-existing scholarship on the particular relationship of prosperity with aid.

The limitation of GDP is further highlighted in the book *Good Economics for Hard Times*, where two winners of the 2019 Nobel Prize in Economics, Abhijit Banerjee and Esther Duflo, point out that a larger GDP doesn't necessarily equate to a rise in human well-being—especially if it isn't distributed equitably—and that the pursuit of it can sometimes be counterproductive. They write that, "nothing in either our theory or the data proves the highest GDP per capita is generally desirable," aligning with the rationale of investigating prosperity as opposed to relying on a purely economic measure.

### 2.1. *What is Prosperity?*

In their book *Beyond GDP*, Joseph E. Stiglitz, Jean-Paul Fitoussi and Martine Durand **(4)** contend that, despite GDP being a powerful economic indicator, it does not provide a holistic review of the health of countries and societies; the measure falls short in indicating the equitable distribution of this growth, environmental sustainability, quality of life, and other factors contributing to the country's success.

These limitations are addressed through the use of prosperity as an alternative measure, a multi-dimensional concept that encompasses economic factors, social well being, and environmental sustainability.

According to the United States Prosperity Index, genuine prosperity extends beyond society's economy or an individual's financial wealth: it represents an environment in which members of society are enabled to meet their full potential. A prosperous society often characterized by having effective institutions, an open economy, and empowered people who are healthy and educated **(5).**

The Legatum Institute **(1)**, known for its extensive work on the issue, offers a comprehensive definition of prosperity that encompasses a variety of factors in a country. These factors range from health, education, and governance to personal freedom, social capital and enterprise conditions, among others. Significantly, prosperity serves as a proxy for an open economy to harness ideas and talent to create sustainable pathways out of poverty.

### 2.2. *Linkage Between Prosperity and Official Development Assistance (ODA)*

According to "One" **(6),** a prominent nonprofit organization dedicated to acting for global justice, ODA ("global aid") denotes the transfer of capital from predominantly more affluent countries to those which are developing; this transfer is primarily aimed at combating poverty and supporting economic development.

In the absence of widespread research on measuring aid effectiveness in increasing prosperity itself, existing literature delves into traditional measures of linking aid and poverty. Alongside humanitarian relief efforts, many aid initiatives aim to generate economic growth and reduce poverty in recipient states. These goals are frequently measured using GDP growth, GDP per capita, or daily income. For example, The World Bank Group, in pursuit of its mission is to end extreme poverty and promote shared prosperity, uses the number of people living under $2.15 per day (who the organization classifies as living in "extreme poverty") as a benchmark to monitor global progress.

Researchers have spent decades debating, without resolution, the cross-country relationship between foreign aid and economic growth. Some studies suggest that aid has a robust positive impact on economic growth, while others fail to distinguish any significant relationship.

In the research paper "Counting Chickens When They Hatch: Timing and the Effects of Aid on Growth" Michael A. Clemens, and Steven Radelet et al. 2011 **(7)**. reanalyze data from the three of the most influential published aid-growth studies, finding that aid causes some growth in recipient countries. However, the magnitude of this relationship is modest, varies greatly across recipients, and diminishes at high levels of aid.

This is contrasted by the findings of R. Glenn Hubbard and William Duggan in their book *The Aid Trap: Hard Truths About Ending Poverty*, 2009 **(8)**, where they argue that most aid, though well-intentioned, actually undermines the development of local business sectors in the poorest nations. The authors assert that developing countries can attain prosperity and self-sufficiency only if aid money goes to cultivating a functioning business sector.

This view on the limitations of foreign aid is expounded upon in William Easterly's book *The White Man's Burden*, where he delves into the shortcomings of foreign aid through two contrasting examples. Easterly begins by describing how the West spent US$2.6T in aid in the last five decades only to fail to get 12-cent malaria medicines and four-dollar bed nets to the children

and low-income families in need of them. In contrast, on July 16, 2005, the American and British economies delivered *nine million* copies of the sixth volume of Harry Potter to eager fans all in a single day **(9)**. This contrast prompts an exploration into the stark difference in efficiency and infrastructure supporting the two efforts, as well as the role of economic incentivization in fostering prosperity.

## 2.3. *ODA Sectors*

A significant proportion of aid is directed toward health and humanitarian relief; however, there are also specific sectors which are targeted to help develop economic growth and prosperity. The study places a particular emphasis on two key sectors within this context: education and economic infrastructure and services.

### 2.3.1. *Education*

In 1957, Nobel Laureate Robert Solow described increasing levels of education as one of the key factors that contributed to the income growth of nations. This idea is supported by the paper "Education and Economic Growth" by Hanushek and Woßman, 2010 which offers a comprehensive synthesis of the theoretical literature on the topic. The paper expands on how education increases the innovative capacity of an economy, promoting the creation and dissemination of new knowledge, technologies, products and processes to foster growth. The paper also provides a valuable review of the impact of the quality and quantity of education on economic growth **(10).**

The positive impact of education on economic growth is further underscored in a recent OECD report, which stated that providing every child with access to education and the skills needed to participate fully in society would boost GDP by an average of 28% per year in lower-income countries and 16% per year in high-income countries for the next 80 years **(11)**. While education has long been suggested as a vital determinant of economic growth, its impact on lasting prosperity creation is mainly discussed in the context of economic growth. The paper expands the analysis to uncover its impact on prosperity.

### 2.3.2. *Economic Infrastructure and Services*

The economic infrastructure and services sector refers to assistance for networks, utilities, and other services that facilitate economic activity. Notable among these are transport and storage, communications, energy generation and distribution efficiency, banking and financial services, and the general promotion of business activity. Research conducted by Caldero´n and Serve´n in their 2004 study "The Effects of Infrastructure Development on Growth and Income Distribution" **(12)** found a positive relationship between growth and stock of infrastructure assets, with income inequality declining with a higher quality of infrastructure.

A similar narrative produced by the five member nations (Australia, Canada, New Zealand, the United Kingdom, and the United States) in 2015 provided a strategic overview of the relationship between critical infrastructure and national prosperity. In their view, a well-designed infrastructure is both an essential driver of national prosperity and a prerequisite for future growth and economic expansion **(13)**; by creating jobs and putting more efficient systems in place, it enhances a nation's productivity, quality of life, and economic progression, ultimately providing the foundational networks on which businesses and individuals rely.

Taking the prior juxtaposition of the distribution of Harry Potter books and malaria medication in The White Man's Burden, These same links between infrastructure investment and economic outcomes are highlighted in the 2021 book Economic Analysis and Infrastructure Investment by Edward L. Glaeser & James M. Poterba, which indicates that public infrastructure spending has minor stimulative effects on output in the short run but significant effects in the long run **(14)**.

## 2.4. *Innovation*

The linkages between innovation and economic growth have been a central theme in the discourse on economic development since the early twentieth century. In his 1921 work, The Theory of Economic Development, Joseph A. Schumpeter writes on the role of innovation and entrepreneurship in economic growth, as does Robert Solow in his 1956 work on the topic **(15, 16)**. This impact at the macroeconomic level is exemplified by the impact of electricity and railroad technology on economies in the late 19th century, as well as those of the communications and computer revolutions of the 1990s.

This idea is further supported by the report "The New Challenge to America's Prosperity: Findings from the Innovation Index", 1999, in which members Michael E. Porter and Scott Stern of the Council on Competitiveness stress the tight linkage between a nation's innovative capacity and prosperity in particular **(17)**.

Delving more specifically into the relationship, the book The Prosperity Paradox, makes the argument that a particular type of innovation generates a long term, sustained economic prosperity **(18)**. A standard definition of innovation describes it as the transformation of knowledge into new products, processes, and services involving more than just science and technology. However, as per the book's framework, there are three distinct types of innovation that yield varying impacts on an economy:

**Sustaining innovations** improve existing products/services and target people who can afford the product or service in a particular segment.

**Efficiency innovations** enable companies to do more with fewer resources.

**Market-creating innovations** create new markets that serve people for whom either no products existed or existing products were inaccessible.

The market creating innovations is a powerful concept as it targets a large population segment in any nation's economy, helping embed itself in an economy through a previously unaddressed market. In his book The Fortune at the Bottom of Pyramid, C. K. Prahlad **(19)** draws attention to the substantial demographic, comprising 4-5 billion individuals, who remain either unserved or underserved by the conventional large-scale organized private sector. This segment, when effectively catered to through the creation of tailored market-creating innovations, can serve as a vital catalyst for global trade and overall prosperity. Research by the International Finance Corporation suggests that this segment accounts for an ample US$5 trillion **(19)**.

### 2.5. *The Rule of Law*

The presence of effective rule of law in a state serves as another critical determinant that facilitates innovation and prosperity. This concept is addressed in Hernando de Soto's book, The Mystery of Capital: Why Capitalism Triumphs in the West and Fails Everywhere Else **(20)** which highlights how assets cannot become capital unless the country guarantees the rule of law. Historian Sven Beckert in his book Empire of Cotton, A Global History **(21)** emphasizes the historical prevalence of this idea, dating as far back to the cotton trade of 1835.

A number of other sources, among these the United Nations Global Compact **(22)**, support this same idea; without proper rule of law, there is no clear ownership of assets and property transfer, which are critical for businesses and, by extension, economies to prosper.

The rule of law concept is multifaceted, comprising discrete components ranging from the security of person and property rights to checks on government and corruption. As explained by the legal origin theory, some countries have legal systems that support more dynamic market economies than others (23). The paper does not attempt to analyze the theory underlying these different causal mechanisms linking the rule of law to economic growth, but rather emphasizes its critical role in building an environment that is conducive to it.

## 3. **METHODOLOGY AND DATA**

### 3.1. *Data*

The study incorporates a selection of relevant variables, including Net ODA Received Per Capita, Legatum Prosperity Score, Innovation Score, Rule of Law, and two ODA sector variables: Education and "Economic and Infrastructure Services."

A dataset was constructed using panel data on 74 world countries from 2013–2021, including only states which had received Official Development Assistance during the period. This study period was selected based on the availability of data for the variables, and is therefore unbalanced where scores were not available for certain years. Data sources include the Legatum Institute, The World Bank, the OECD, and the Global Innovation Index (GII).

#### 3.1.1. *Dependent Variable*

- **Legatum Prosperity Index (i)**: The Legatum Prosperity Index (Score) is the dependent variable of the study; it serves as an indicator of the wealth and well-being of countries. Within the context of the study, it is used as a proxy for a nation's prosperity. Some literature suggests GDP growth as a symbol of prosperity, but it lacks other elements such as health, Education, Governance, personal freedom, enterprise conditions, and others. The raw score was utilized instead of the index to include continuous numbers in the modeling. The index assesses the performance of 167 countries across 12 pillars to provide a holistic picture of prosperity, accounting for measures such as health, education, governance, etc.

#### 3.1.2. *Independent Variables*

- **Net ODA Received Per Capita (ii)**: The independent variable net ODA received per capita (current US$) is calculated by dividing net official development assistance received by the mid-year population estimate of the recipient state. The calculations adhere to the guidelines established by the Development Assistance Committee (DAC) **(24)**.

- **Innovation Score (iii)**: Innovation is at the centerpiece of many governments' strategies for economic growth. The innovation score, obtained from the Global Innovation Project's Global Innovation Index (GII), ranks world economies according to their innovation capabilities; this ranking process consists of roughly 81 indicators, among these market sophistication, human capital, and creative and knowledge output, grouped into innovation inputs and outputs. The GII aims to capture the multi-dimensional nature of innovation.

- **Rule of Law Score (iii)**: The rule of law is another independent variable captured by the Global Innovation Index (GII). The metric captures perceptions of the extent to which agents have confidence in and abide by the rules of society, encompassing the quality of contract enforcement, property rights, the effectiveness of police and the courts, and the likelihood of crime and violence.

- **ODA Education Sector (ii)**: The ODA Education Sector metric encompasses a range of identifiable general education components: these span from policy and administration to provisions at primary, secondary, and tertiary levels (including multi-sector and vocational training), to aspects of culture and recreation **(25)**.

- The OECD's work on education helps individuals and nations identify and develop the knowledge and skills that drive better jobs and lives, generate prosperity, and promote social inclusion **(26)**.

- **ODA Economic Infrastructure and Services (ii)**: In the DAC sectoral classification, economic infrastructure and services relate to assistance for networks, utilities, and services that facilitate economic activity. Notably among these are transport and storage, communications, energy generation, distribution efficiency, banking and financial services and other business services.

## 3.2. *Mathematical Model*

The study employs five regression models to examine the impact of the independent variables on the prosperity score. All variables have been naturalized using the natural logarithm function to account for heteroskedasticity (unequal scatter of residuals).

**Equation 1**:

The first equation uses a linear model to study the relationship between Net ODA Received Per Capita and prosperity score. The hypothesis is that net ODA received has a negligible to weak correlation with a nation's prosperity.

$$lnpp_{i,t} = a + b_1 lnodapc_{i,t-1} + u_i + e_i$$

**Equation 2**:

The second equation uses a linear model to study the relationship between the innovation and prosperity scores. The hypothesis is that innovation has a moderate to strong correlation with nations' prosperity.

$$lnpp_{i,t} = a + b_1 lnis_{i,t-1} + u_i + e_i$$

**Equation 3**:

The third equation uses a linear model to study the impact of the rule of law on the prosperity score. The hypothesis is that the rule of law positively affects the innovation environment and therefore has a positive relationship with a nations' prosperity.

$$lnpp_{i,t} = a + b_1 lnrl_{i,t-1} + u_i + e_i$$

**Equations 4 and 5**:

Lastly, the fourth and fifth equations use a linear model to study the impact of two ODA receiving sectors, education and economic and infrastructure services, on the prosperity score. The hypothesis is that, as net ODA will likely exhibit a weak relationship with prosperity, these two measures will do the same.

$$lnpp_{i,t} = a + b_1 lnaideconinfra_{i,t-1} + u_i + e_i$$
$$lnpp_{i,t} = a + b_1 lnaidedu_{i,t-1} + u_i + e_i$$

**Where**:
**Dependent Variable**
LNPP = prosperity score

**Independent Variables**
LNODAPC = Net ODA Received Per Capita
LNRL = Rule of Law Score

LNIS = Innovation Score

*Independent Aid Distribution Sector Variables*
LNAIDEDU = Aid distribution to the education sector.
LNAIDECONINFRA = Aid distribution to economic Infrastructure and services.

$a, b_1$ = Estimation parameters.
$u_i$, $e_i$ are unobserved and observed errors.
 The subscript i shows countries included in the model (i=1, 2, 3, … 74), while the subscript t indicates the time period (2013–2021).

### 3.3. *Estimation Approach*

 A series of statistical tests were applied to first confirm the validity of the econometric approach, then to determine the relationships between the variables:

- **Descriptive statistics:** Used to describe and understand the data set's features through broad observations about the samples and measures.

- **Pairwise Pearson Correlation:** Measures the strictly linear correlation between variables in the data set. It shows whether changes in one variable are associated with changes in other variables **(27)**.

- **Panel unit root test:** Confirms whether the time series used in the study is stationary and no seasonal patterns exist. The null hypothesis is that the series is non-stationary; we reject the null hypothesis if the p-value is less than 5% **(28)**.

- **Hausman test:** Determines whether a fixed effect or random effect model is appropriate for the analysis. This is done to ensure that random effects correlated with the explanatory variables are accounted for. The null hypothesis is that the preferred approach is the random effect model **(29)**.

- **FMOLS, Fully Modified Ordinary Least Squares,** is an efficient estimation method developed by *Phillips and Hansen in 1995* **(30)** to estimate the long-run relationship between two or more variables, all the while accounting for cross-sectional heterogeneity and endogeneity in the panel data set. It is used to provide optimal estimates of cointegrating regressions. The method modifies least squares to account for serial correlation effects and for the endogeneity in the regressors resulting from a cointegrating relationship.

- **GMM, a Generalized Method of Moments**, is useful with dynamic panel data with fixed effects. Developed by *Arellano and Bond 1991* **(31) (32)**, it is another model that accounts for endogeneity and serial correlation issues. The GMM combines observed economic data with information on population moment conditions to produce estimates of the unknown parameters of the economic models or the equations. The model is feasible to use in the study as the cross-sectional dimensions (74 countries) are larger than the time dimension (2013-2021).

Both FMOLS and GMM are used to compare and validate results for consistency.

## 4. **RESULTS AND DISCUSSION**

### 4.1. *Descriptive Statistics*

**Table 1** provides the results of descriptive analysis based on the 633 observations.

As per the results, variables LNODAPC, LNRL, and LNAIDEDU have a kurtosis value greater than three, exhibiting a leptokurtic distribution. This means the distributions have wider tails than a normally distributed dataset, indicating outliers and/or extreme values. On the other hand, LNPP, LNIS, and LNAIDECONINFRA have a kurtosis value of less than three and display a platykurtic distribution, indicating less significant outliers.

The Jarques-Bera test confirms that none of the variables are normally distributed at a 5% significance level; the lack of normality is of little concern in FMOLS and GMM regression models.

| Summary | LNPP | LNODAPC | LNRL | LNIS | LNAIDEDU | LNAIDECONINFRA |
|---|---|---|---|---|---|---|
| Mean | 3.961391 | 3.220163 | 3.416325 | 3.338743 | 3.744270 | 3.553497 |
| Median | 3.994524 | 3.617769 | 3.487375 | 3.367296 | 3.819908 | 3.807329 |
| Maximum | 4.241327 | 5.798990 | 4.432007 | 3.848018 | 6.477111 | 8.738469 |
| Minimum | 3.484312 | -3.947031 | 1.193922 | 2.610070 | 0.148420 | -4.605170 |
| Std. Dev. | 0.146219 | 1.483374 | 0.476490 | 0.216506 | 1.092744 | 2.167365 |
| Skewness | -0.517344 | -1.322848 | -0.902952 | -0.461788 | -0.475326 | -0.385532 |
| Kurtosis | 2.859780 | 5.678379 | 4.359087 | 2.878564 | 3.165511 | 2.861981 |
| Jarque-Bera | 28.75509 | 373.8240 | 134.7342 | 22.88665 | 24.55865 | 16.18340 |
| Probability | 0.000001 | 0.000000 | 0.000000 | 0.000011 | 0.000005 | 0.000306 |
| Sum | 2507.560 | 2038.363 | 2162.533 | 2113.424 | 2370.123 | 2249.364 |
| Sum Sq. Dev. | 13.51222 | 1390.651 | 143.4907 | 29.62499 | 754.6643 | 2968.802 |
| Observations | 633 | 633 | 633 | 633 | 633 | 633 |

**Table 1.**

### 4.2. *Correlation Analysis*

**Table 2** provides the results of Pearson's correlation analysis.

A strong positive correlation is observed between the prosperity score (LNPP) and innovation score (LNIS), as well as between the prosperity score and rule of law (LNRL). This indicates that a higher value for each of these variables is associated with a higher prosperity score.

Net ODA per capita (LNODAC), aid distribution to economic infrastructure and services development (LNAIDE- CON INFRA), and aid distribution to the education sector (LNEDU) exhibit a low to moderate negative correlation with LNPP. This suggests that higher levels of these variables are associated with lower prosperity scores, and may be owed to reverse causality (i.e. less prosperous nations simply requiring more development assistance).

| | LNPP | LNODAPC | LNINS | LNRL | LNAIDECONINFRA | LNAIDEDU |
|---|---|---|---|---|---|---|
| LNPP | 1 | -0.233043 | 0.787896 | 0.668736 | -0.098277 | -0.341685 |
| LNODAPC | -0.233043 | 1 | -0.26543 | -0.047321 | 0.09862 | -0.020019 |
| LNINS | 0.787896 | -0.26543 | 1 | 0.567176 | 0.01454 | -0.133188 |
| LNRL | 0.668736 | -0.047321 | 0.567176 | 1 | 0.065622 | -0.131353 |
| LNAIDECONINFRA | -0.098277 | 0.09862 | 0.01454 | 0.065622 | 1 | 0.518769 |
| LNAIDEDU | -0.341685 | -0.020019 | -0.133188 | -0.131353 | 0.518769 | 1 |

**Table 2.**

### 4.3. *Panel Unit Root Test*

**Table 3** displays the results of the unit root tests. All variables have a unit root at the level or first difference, thereby rejecting the null hypothesis at the 5% significance level. According to the findings of these tests, the variables do not show seasonal variations, implying that they have stable means, variances, etc. over time. This further indicates that the variables are appropriate for further analysis using a co-integration test (FMOLS).

| Variable | Fisher-PP | | Fisher-ADF | | IPS | | LLC | |
|---|---|---|---|---|---|---|---|---|
| | I(0) | I(1) | I(0) | I(1) | I(0) | I(1) | I(0) | I(1) |
| | | | | | | | | |
| LNPP | 200.51 | 360.693 | 100.075 | 217.079 | 2.58318 | -3.97197 | -5.88386 | -20.43 |
| | (0.0002) | (0.000) | (0.9874) | (0.000) | (0.9951) | (0.000) | (0.000) | (0.000) |
| | | | | | | | | |
| LNODAPC | 287.459 | 578.365 | 147.662 | 285.066 | -0.17815 | -10.6083 | -1.53185 | -74.4315 |
| | (0.1982) | (0.000) | (0.000) | (0.000) | (0.4293) | (0.000) | (0.0628) | (0.000) |
| | | | | | | | | |
| LNINS | 170.96 | 633.522 | 95.3157 | 206.42 | 3.52197 | -3.21073 | -2.68941 | -11.9264 |
| | (0.0171) | (0.000) | (0.9953) | 0.0001 | (0.9998) | (0.0007) | (0.0036) | (0.000) |
| | | | | | | | | |
| LNRL | 260.644 | 682.838 | 147.615 | 248.29 | -1.49149 | -5.0961 | -11.4008 | -20.9882 |
| | (0.000) | (0.000) | (0.1989) | (0.000) | (0.0679) | (0.000) | (0.000) | (0.000) |
| | | | | | | | | |
| LNAIDECONINFRA | 370.289 | 688.488 | 210.404 | 302.563 | -3.52688 | -7.722689 | -11.8387 | -21.0998 |
| | (0.000) | (0.000) | (0.000) | (0.000) | (0.0002) | (0.000) | (0.000) | (0.000) |
| | | | | | | | | |
| LNAIDEDU | 298.727 | 625.995 | 156.029 | 302.463 | -1.1809 | -8.25648 | -8.1164 | 23.008 |
| | (0.000) | (0.000) | (0.0937) | (0.000) | (0.1188) | (0.000) | (0.000) | (0.000) |

**Table 3.** p-values are in the parentheses

### 4.4. *Hausman Test*

The results of the Hausman test, as displayed in **Table 4**, indicate that all of the variables have p-values less than 0.05. This allows us to reject the null hypothesis, which states that a random effect model is suitable for the analysis, and instead accept the fixed effect model. This helps account for variables that were not included in the estimation that could have had an impact on the dependent variable (omitted variable bias).

| Variable | Fixed | Random | p value |
|---|---|---|---|
| | | | |
| LNODAPC | 0.00542 | -0.001022 | 0.0000 |
| | | | |
| LNINS | -0.0377266 | -0.005846 | 0.0000 |
| | | | |
| LNRL | 0.040291 | -0.054513 | 0.0000 |
| | | | |
| LNAIDECONINFRA | 0.000366 | 0.000186 | 0.0115 |
| | | | |
| LNAIDEDU | 0.002431 | 0.000365 | 0.0000 |

**Table 4.**

### 4.5. *Generalized Method of Moments (GMM) Results*

**Table 5** provides the results of the GMM estimation, summarizing the estimation results of the four equations specified above. The model uses levels of the dependent variable lagged one level.

| Variable | Equation 1 | Equation 2 | Equation 3 | Equation 4 | Equation 5 |
|---|---|---|---|---|---|
| LNPP (p-value) | 0.0000* | 0.0000* | 0.0000* | 0.0000* | 0.0000* |
| Coefficient | 0.949761 | 0.954172 | 0.919743 | 0.981512 | 0.970362 |
| LNODAPC (p-value) | 0.9001** | | | | |
| Coefficient | 0.000135 | | | | |
| LNINS (p-value) | | 0.6641** | | | |
| Coefficient | | -0.004513 | | | |
| LNRL (p-value) | | | 0* | | |
| Coefficient | | | 0.00585 | | |
| LNAIDECONINFRA (p-value) | | | | 0.6426** | |
| Coefficient | | | | -0.000277 | |
| LNAIDEDU (p-value) | | | | | 0.0186* |
| Coefficient | | | | | 0.001393 |
| | | | | | |
| Probability (J-Statistic) | 0.225492 | 0.119733 | 0.315261 | 0.315261 | 0.332475 |
| | | | | | |
| Periods Included | 7 | 7 | 7 | 7 | 7 |
| | | | | | |
| Cross Section | 74 | 74 | 74 | 74 | 74 |
| | | | | | |
| Total Panel Observations | 487 | 487 | 487 | 487 | 487 |

**Table 5.** Dynamic Panel GMM Results (**P>0.05, *P<0.05)

Because the p-value acquired from the J-statistic is greater than 0.05, we fail to reject the null hypothesis and can conclude that the GMM is a good fit for the data.

1. The output reveals a positive coefficient for the relationship between LNODAPC (net ODA received per capita) and the LNPP (prosperity score), the dependent variable. However, because the p-value is significantly higher than 0.05, we fail to conclude that there is any statistically significant relationship between net ODA received per capita and prosperity score.

2. LNIS (Innovation score) exhibits a negative coefficient, with the p-value significantly higher than 0.05. This indicates that there is no statistically significant relationship between the innovation and prosperity score in the dataset.

3. LNRL (Rule of law) shows a positive coefficient, with a p-value of less than 0.05. This is indicative of a positive, statistically significant relationship with the dependent variable. According to the results, a higher rule of law positively affects prosperity.

4. LNAIDECONINFRA (Economic Infrastructure and Services) shows a negative coefficient, with a p-value is significantly higher than 0.05. This means there is no statistically significant relationship between foreign aid specifically targeted toward economic infrastructure and services and the prosperity score.

5. LNAIDEDU (Education Sector) displays a positive coefficient, with a p-value of less than 0.05. This is indicative of a statistically significant relationship with the dependent variable. According to the results, increased foreign aid targeted toward the education sector positively impacts prosperity.

#### 4.6. *Fully Modified Ordinary Least Squares (FMOLS) Regression Results*

**Table 6** provides the results of the FMOLS regression. It summarizes reports from the estimation results of the four equations specified above. The model uses levels of the dependent variable lagged one level. .

| Variable | Equation 1 | Equation 2 | Equation 3 | Equation 4 | Equation 5 |
|---|---|---|---|---|---|
| LNPP (p-value) | 0.0000* | 0.0000* | 0.0000* | 0.0000* | 0.0000* |
| Coefficient | 0.935647 | 0.928034 | 0.910627 | 0.93769 | 0.934963 |
| LNODAPC (p-value) | 0.8524** | | | | |
| Coefficient | 0.000131 | | | | |
| LNINS (p-value) | | 0.0474* | | | |
| Coefficient | | -0.011041 | | | |
| LNRL (p-value) | | | 0.0001* | | |
| Coefficient | | | 0.010016 | | |
| LNAIDECONINFRA (p-value) | | | | 0.7144 | |
| Coefficient | | | | 0.000125 | |
| LNAIDEDU (p-value) | | | | | 0.2113** |
| Coefficient | | | | | 0.001532 |
| | | | | | |
| R squared | 0.996236 | 0.996272 | 0.996379 | 0.996233 | 0.996248 |
| | | | | | |
| Adjusted R squared | 0.995549 | 0.995591 | 0.995713 | 0.995546 | 0.995564 |
| | | | | | |
| Periods Included | 7 | 7 | 7 | 7 | 7 |
| | | | | | |
| Cross Section | 74 | 74 | 74 | 74 | 74 |
| | | | | | |
| Total Panel Observations | 487 | 487 | 487 | 487 | 487 |

**Table 6.** Fully Modified Ordinary Least Squares results (**P>0.05, *P<0.05)

1. The output displays a positive coefficient for the relationship between LNODAPC (net ODA received per capita) and LNPP (prosperity score); however, because the p-value is significantly higher than 0.05, we fail to determine any statistically significant relationship between aid and prosperity.

2. LNIS (Innovation score) exhibits a negative coefficient, with a p-value of less than 0.05. This indicates that the dataset shows a statistically significant negative relationship between the innovation and prosperity scores at a 5% significance level.

3. LNRL (Rule of law) shows a positive coefficient, with a p-value of less than 0.05. This is indicative of a positive, statistically significant relationship with the dependent variable. According to the results, a higher rule of law positively affects prosperity. This value is statistically significant up to the .01% significance level.

4. LNAIDECONINFRA (Economic Infrastructure and Services) shows a negative coefficient, with a p-value is significantly higher than 0.05. This means there is no statistically significant relationship between foreign aid specifically targeted toward economic infrastructure and services and the prosperity score.

5. LNAIDEDU (Education Sector) displays a positive coefficient with a p-value greater than 0.05. This means that, according to the FMOLS, we fail to find a statistically significant relationship between aid directed toward the education sector and prosperity score.

According to the results displayed in **Table 6**, two of the five statistics refute the null hypothesis of no cointegration at a 5% significance level. In summary, we conclude from the cointegration test that there is strong empirical evidence that the prosperity score has a long-running positive association with the rule of law and a negative association with innovation in developing countries from 2013 to 2021.

The R squared value shows that the model is a very strong fit for the data, indicating that 99% of the variability in results is accounted for by the model.

The R-squared value shows that the model perfectly fits the data. It indicates that 99% of the variance in the LNPP that the independent variables explain collectively.

### 4.7. *Summary Results Comparison*

A summary of both the GMM and FMOLS results is below for a quick comparison. Three of the five results are consistent in statistical significance between the GMM and FMOLS regression, with four of the five being consistent in terms of coefficient sign.

| Variable | GMM | FMOLS |
|---|---|---|
|  | (**P>0.05, *P<0.05) |  |
|  |  |  |
| LNODAPC | (0.9001)** | (0.8524**) |
|  | 0.000135 | 0.000131 |
|  |  |  |
| LNINS | (0.6641**) | (0.0474*) |
|  | -0.004513 | -0.011041 |
|  |  |  |
| LNRL | (0.0000*) | (0.0001*) |
|  | 0.00585 | 0.010016 |
|  |  |  |
| LNAIDECONINFRA | (0.6426**) | (0.7144**) |
|  | -0.000277 | 0.000125 |
|  |  |  |
| LNAIDEDU | (0.0186*) | (0.2113**) |
|  | 0.001393 | 0.001532 |

**Table 7.**

## 5. **INTERPRETATION**

### 5.1. *Official Development Assistance*

The attached statistical analysis of data between 2013 and 2021 for aid-receiving countries shows a negative correlation between ODA received per capita and prosperity score. More robust models, such as the GMM and FMOLS, show that ODA received per capita has a statistically insignificant relationship with the prosperity score; this supports the hypothesis that aid does not move the needle in creating lasting prosperity.

### 5.2. *Innovation*

The initial Pearson correlation analysis reveals a positive relationship between the innovation and prosperity scores. However, further scrutiny with the GMM and FMOLS regression instead shows a statistically significant negative relationship between the two variables.

The result seems counterintuitive to the argument that investments in innovations create lasting prosperity. However, this insight can be explained through Professor Christensen's classification of the distinct types of innovation. As discussed earlier, the unique nature of market-creating innovations make them difficult to reliably pinpoint and measure, making them difficult to account for with innovation indices such as the ones utilized in this paper.

The analysis underscores the potential for certain types of innovation to function against long-term growth. For example, resource-rich nations such as Nigeria or Saudi Arabia often employ efficiency innovations to cut costs and increase profit margins. This process can inadvertently function against prosperity by cutting jobs and promoting a subsequent reduction in economic activity. Examining innovations on a case-by-case basis can help further understand the potential of market-creating innovations, in particular, in stimulating prosperity.

### Reliance Jio Under the Lens of Market Creating Innovations

Reliance Jio is a prominent story of an innovation that addressed a previously underserved market, helping India expand its telecommunications infrastructure to hundreds of millions of individuals across the country. By democratizing access to an array of essential services—services encompassing entertainment, commerce, communication, healthcare, education, and more—Jio has effectively empowered a significant segment of the population and unlocked substantial growth potential in the process.

According to Reliance Industries Chairman Mukesh Ambani, the idea of Jio was seeded by his daughter: "She was a student at Yale (in the US) and was home in Mumbai, India, for the holidays. She wanted to submit some coursework and said, "Dad, the internet in our house sucks" **(33)**. At the time, India suffered from poor connectivity and a severe scarcity of its most critical digital resource: data. According to Ambani, data was not only scarce, but it was priced artificially high to make it unaffordable to a majority of Indians. Most of India's population, whose average income was less than $150 per month, had never had internet access.

In 2016, however, Jio was launched and things changed; the company offered its services at rock-bottom prices, coupling these with extended introductory offers that made the product affordable to the large, formerly unaddressed population. Soon after its launch, Jio acquired over 425 million subscribers (34). Millions of working-class and rural Indians could video call relatives, stream cricket matches, or play online games for the first time.

Jio targeted customers like 59-year-old potato farmer Govind Singh Panwar. His home in the Himalayan foothills was built of mud and stone, with his village having no paved roads or indoor plumbing. Still, he had broadband internet access **(33)**.

To make this happen, Jio was pushed to build extensive telecommunications infrastructure, distribution channels, and marketing strategies that reached a large swath of the previously unreachable population. The telecom network covered 18,000 cities and towns and 200,000 villages, touching many without electricity. This effort required more than 200,000 cell towers and 150,000 miles of high-tech fiber-optic cable, which, according to the company, were enough to circle the Earth six times **(33)**.

To protect its sprawling web of cables, Reliance also hired a national network of ex-army staff and local residents to look after its lines. By "creating" this previously unaddressed market, Jio has embedded itself in the economy, responsible for a number of impressive figures:

1. Today, with an annual revenue of $10 billion, Jio employs more than 90,000 employees, providing them 14,349,839 person-hours of training **(35)**.

2. The ecosystem has grown, and the company's marketplace now offers education, healthcare, entertainment, banking, and manufacturing services.

3. Jio connects over 50 million micro, small, and medium businesses (MSMB) through its services, streamlining coordination among employees and supply chain partners while connecting with other MSMBs and customers on the platform. Jio also offers improved access to banks, commerce, suppliers, and merchants **(36)**.

4. The platform has attracted more than $20 billion in investment from Facebook parent Meta Platforms, Alphabet, Intel, and other investors (37).

5. Jio's corporate social responsibility initiatives have served over 45 million people in 44,700 villages **(37)**.

6. Jio's network carries almost 10% of the global mobile data traffic **(37)**.

7. According to the Institute for Competitiveness, a calculation conducted prior to Jio's entry into India's telecommunications industry found that the company's doing so would expand India's per capita GDP by about 5.65% **(38)**.

Jio's groundbreaking initiatives have democratized access to previously exclusive services, cultivating new employment opportunities and customers as a foundation for India's pursuit of greater prosperity.

### 5.3. *Rule of Law*

The data between 2013 and 2021 for the aid-receiving countries shows a positive correlation between the rule of law and the innovation and prosperity scores. This is supported by the GMM and FMOLS models, which display a positive, statistically significant coefficient between the rule of law and prosperity and innovation scores. This supports the hypothesis that rule of law positively impacts the prosperity of nations.

Laws are often frequently preceded by innovation due to a need to govern the implications of innovation. This is illustrated by regulations following the recent surge in artificial intelligence (AI), such as the European Union's 2023 Artificial Intelligence Acts; the legislation accounts for a number of recent facets of the technology, such as transparency in the training of generative AI models **(39)**.

### 5.4. *ODA Sectors*

#### 5.4.1. *Economic Infrastructure and Services*

The data between 2013 and 2021 for aid-receiving countries shows a negative correlation between aid given to the economic infrastructure and services sector and prosperity score. More robust models, such as the GMM and FMOLS, show a negative and statistically insignificant relationship between the two variables.

This result appears counterintuitive, supporting that aid put toward building infrastructure to support a nation does not have long-term impacts on a state's prosperity. This may be owed to the fact that, unlike innovations, infrastructure built purely on aid is not self-sustained or guaranteed to be used and maintained. This is a topic that warrants further exploration to understand the factors underlying the lack of a significant relationship between the two variables.

#### 5.4.2. *Education*

The data between 2013 and 2021 for aid-receiving countries reveals a positive relationship between prosperity scores and aid directed toward the education sector. This is corroborated by the GMM, which shows a positive, statistically significant relationship between the two. This goes against the results of both general ODA and ODA directed toward infrastructure, being the only one of the three to have a statistically significant relationship with prosperity. This is in line with the existing literature on the topic; as noted above, education is frequently noted as a driver of new knowledge, increased societal contributions, and innovation within a country.

## 6. CONCLUSION & POLICY IMPLICATIONS

The common wisdom that alleviating poverty through aid will create lasting prosperity differs from what we have seen in the literature and statistical analysis and is supported by the results of the study.

Through an econometric examination of the dataset, the study determined that there is not a statistically significant relationship between net ODA per capita and the prosperity score of the included nations in the given time period. The same was found to be true for the relationship of the innovation score and prosperity, with this metric potentially being owed to variances between types of innovations. The next logical step in this process would be to expand the study to include the impacts of venture capital deal activity on an area, accounting for the various sectors in which investments were targeted.

The relationships of rule of law and ODA directed toward education were found to be positive, supporting notions from existing literature of the two being foundational aspects upon which prosperity thrives. As per existing literature on the topic, education and rule of law, respectively, were also found to be factors that frequently preceded and succeeded innovation.

On the contrary, aid directed toward infrastructure and services on its own was found to have a negligible relationship with prosperity; this may offer support for the theory surrounding the need of "self-sustaining" additions to a country's economy—such as those added through particular types of innovation—to have a long-term impacts on the prosperity of a state. This is because aid on its own does not always have a clear path to cycle through an economy.

This brings us back to the question previously raised by *The White Man's Burden*: the question of why the Harry Potter books were delivered to consumers with efficiency, but not the much-needed malaria medications.

The study highlights the need for incentivization in driving private enterprises to establish infrastructure. Just as aid on its own typically does not have any reason to cycle through an economy, there wasn't an economic incentive to develop an ecosystem for suppliers, distributors, and marketers to participate in the delivery of the medication. On the contrary, private enterprise had economic incentives to build the infrastructure necessary to deliver books. Similar to the example covered by Jio, the desire to leverage resources and gain market coverage resulted in the creation of infrastructure to drive resource distribution. That is what we need for the efficient delivery of medicines.

## 7. REFERENCES AND DATA SOURCES